\documentclass[aps,prl,superscriptaddress,preprint]{revtex4-1}

	\usepackage{amsmath}
	\usepackage{amssymb} 
	\usepackage{color}
	\usepackage{makeidx}
	\usepackage{amsfonts}
	\usepackage[ansinew]{inputenc}
	\usepackage[usenames,dvipsnames]{pstricks}
	\usepackage{subfigure}
	\usepackage{epsfig}
	\usepackage[colorlinks,hyperindex]{hyperref}

	\hypersetup
	{
		colorlinks,%
		citecolor=black,%
		linkcolor=black,%
		urlcolor=black,%
	}

\makeindex
\begin{document}

\title{Network of thermoelectric nanogenerators for low power energy harvesting}

\author{Dimitri Tainoff}
\affiliation{Institut N\'EEL, CNRS, 25 avenue des Martyrs, F-38042 Grenoble, France}
\affiliation{Univ. Grenoble Alpes, Inst NEEL, F-38042 Grenoble, France}

\author{Ana\"{\i}s Proudhom}
\affiliation{Institut N\'EEL, CNRS, 25 avenue des Martyrs, F-38042 Grenoble, France}
\affiliation{Univ. Grenoble Alpes, Inst NEEL, F-38042 Grenoble, France}

\author{C\'eline Tur}
\affiliation{Institut N\'EEL, CNRS, 25 avenue des Martyrs, F-38042 Grenoble, France}
\affiliation{Univ. Grenoble Alpes, Inst NEEL, F-38042 Grenoble, France}

\author{Thierry Crozes}
\affiliation{Institut N\'EEL, CNRS, 25 avenue des Martyrs, F-38042 Grenoble, France}
\affiliation{Univ. Grenoble Alpes, Inst NEEL, F-38042 Grenoble, France}

\author{S\'ebastien Dufresnes}
\affiliation{Institut N\'EEL, CNRS, 25 avenue des Martyrs, F-38042 Grenoble, France}
\affiliation{Univ. Grenoble Alpes, Inst NEEL, F-38042 Grenoble, France}

\author{Sylvain Dumont}
\affiliation{Institut N\'EEL, CNRS, 25 avenue des Martyrs, F-38042 Grenoble, France}
\affiliation{Univ. Grenoble Alpes, Inst NEEL, F-38042 Grenoble, France}

\author{Daniel Bourgault}
\affiliation{Institut N\'EEL, CNRS, 25 avenue des Martyrs, F-38042 Grenoble, France}
\affiliation{Univ. Grenoble Alpes, Inst NEEL, F-38042 Grenoble, France}

\author{Olivier Bourgeois}
\affiliation{Institut N\'EEL, CNRS, 25 avenue des Martyrs, F-38042 Grenoble, France}
\affiliation{Univ. Grenoble Alpes, Inst NEEL, F-38042 Grenoble, France}

\date{\today}

\begin{abstract}
We report the design, elaboration and measurements of an innovative planar thermoelectric (TE) devices made of a large array of small mechanically suspended nanogenerators (nanoTEG). The miniaturized TE generators based on SiN membranes are arranged in series and/or in parallel depending on the expected final resistance adapted to the one of the load. The microstructuration allows, at the same time, a high thermal insulation of the membrane from the silicon frame and high thermal coupling to its environment (surrounding air, radiations). We show a ratio of 60\% between the measured effective temperature of the membrane, (and hence of the TE junctions), and the available temperature of the heat source (air). The thermal gradient generated across the TE junction reaches a value as high as 60 kelvin per mm. Energy harvesting with this planar TE module is demonstrated through the collected voltage on the TE junctions when a temperature gradient is applied, showing a harvested power on the order of 0.3 $\mathrm{\mu}$Watt for a 1~cm$^2$ chip for an effective temperature gradient of 10~K. The optimization of nanoTEGs performances will increase the power harvested significantly and permit to send a signal by a regular communication protocol and feed basic functions like temperature measurement or airflow sensing.

\end{abstract}


\maketitle

\subsection*{Introduction} 

Wireless sensor network (WSN) are an association of various system being able to measure and transfer information over a network without requiring human intervention. Focusing on the sensor part, the major problem preventing the massive spread of WSN is their power supply. Indeed depending on their location, sensors cannot be wired to the grid and/or the maintenance costs induced by the battery change are not acceptable. The smartest solution to get over this problem is to harvest energy from the environment of sensors in order to make them autonomous. 

Since the environment of sensors are changing drastically depending on their final use, energy harvesting solutions that have been developed until now are devoted to one or very few use cases. Each of these technologies has their own advantages and drawbacks which have been reviewed recently \cite{revueEH1,revueEH2}. However, it is difficult to find a general solution which could be sufficiently versatile to cover lots of WSN use cases \cite{revueEHWSN}. In that context, thermoelectricity has a great potential because thermal gradients are present all around even intermittently, being most of the time a byproduct of other energies coming from mechanical motion, electromagnetic dissipation, or chemical reaction... Despite this advantage, the use of thermoelectric generator (TEG), developed since more than fifty years, is still limited to niche markets \cite{infinergia,leblanc2014}. This is due to numerous reasons among which one finds moderate efficiency, the use of rare materials and, above all, the bulk nature of thermoelectric (TE) module. 

With the emergence of low power communication protocols dedicated to the WSN, there is now a real demand for micro-sources of energy that can deliver power on the order of 100~$\mathrm \mu$Watt. That range of power can be obtained from standard TE module with a temperature gradient of few degrees under stationary operating conditions. However, these conditions of use are not necessarily adapted to small wireless sensors especially when the energy source is intermittent and/or the available volume too limited. This has given a new direction for research that is clearly oriented towards miniaturization of TE modules \cite{PublimicroTEG1, PublimicroTEG2}. One of the first miniaturized TEG was based on the structuration of micro-pillars of TE materials, it allowed to strongly limit the volume of the modules \cite{micropeltsite}. However, the need to evacuate the heat flow for preserving the thermal gradient has led to the use of large heat sink making the whole device quite voluminous. The use of such a massive heat sink is the direct consequence of the 3D architecture of the miniaturized TEG and particularly the fact that the thermal coupling of the face in contact with ambient air is not optimized for convection.

Consequently, a growing interest has emerged regarding new designs of nanoTEG using planar and suspended configurations, essentially thanks to pioneering works using microfabrication technologies \cite{huesgen2008,davila2012,perez2014,yuan2015,nakamura2015,sakata2015,Pennelli2014}. The major advantage of these suspended planar configurations is that the free standing part is well coupled to the surrounding air by convection permitting to the device to work without heat sink. However, at least one major issue remains to be solved before using that kind of module as low power generator. Indeed, using TE modules based on micropatterned membrane leads to designs involving very thin films of TE materials and finally to a sizable internal electrical resistance. Since standard low power electronic is optimized for low electrical impedance energy source, the actual solutions exhibiting internal resistance of more than 10~k$\Omega$ are not well adapted. To summarize, the major challenges for a planar microscaled TE module to be an efficient power source is to create a significant thermal gradient upon small distance, to have low electrical impedance and to collect thermal energy at the micro and nanoscale through a thermoelectric voltage.

Here we report the design, the elaboration and the measurement of innovative planar thermoelectric devices made of a large array of suspended nano-thermoelectric generators (single cell nanoTEG). These nanoTEGs are small enough to be duplicated per hundreds upon centimeter square surface and arranged in series and/or in parallel depending on the expected final resistance adapted to the load. The resistance of the measured devices is on the order of 10$^3$~k$\Omega$ in the studied configuration, opening the door to massive parallel wiring of nanoTEG which could lead to resistance of the order of few tens of ohms. We demonstrates a harvested power on the order of 0.3~$\mathrm{\mu}$Watt for a 1~cm$^2$ chip for an effective temperature gradient of 10~K. Under optimized performances, they can produce enough power (below 100~$\rm \mu$Watt) to send a signal using common communication protocols and feed basic functions, for instance temperature sensing or air flow measurements.

\subsection*{Design and realization.} 

\begin{figure}
	\begin{center}
		\includegraphics[width=10cm]{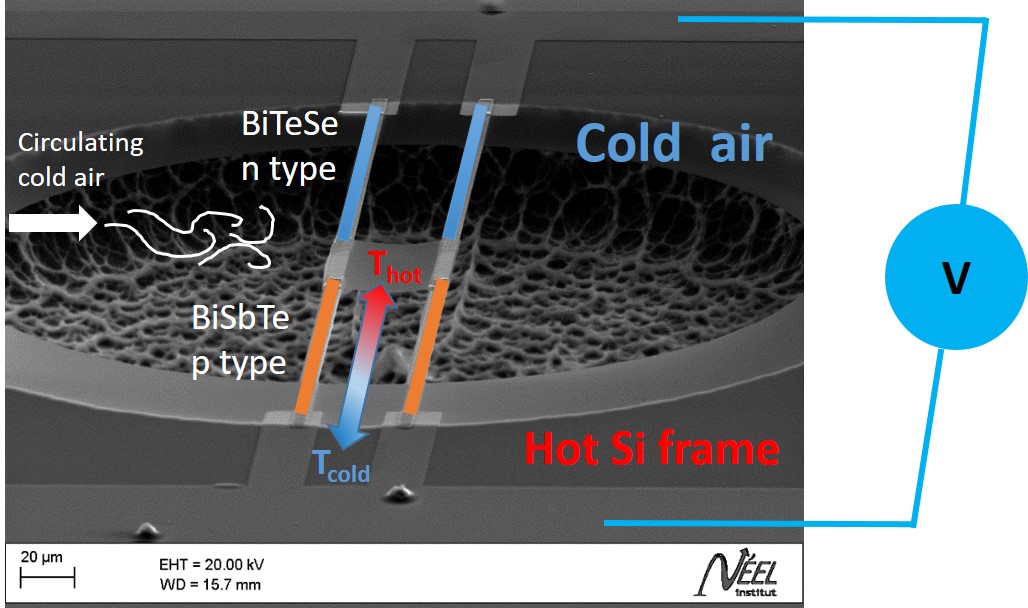}
	\end{center}
	\caption{Scanning Electron Microscope (SEM) image and functioning scheme of a nanoTEG cell composed of a single membrane  (SiN membrane suspended by four SiN arms). Each SiN arm supports a n or p-type Bi$_{2}$XTe$_3$ thin films: the p-type (orange) is Bi$_{2-x}$Sb$_x$Te$_3$ and the n-type (blue) is Bi$_{2-x}$Te$_3$Se$_x$. One cold (hot) thermoelectric junction is located on the membrane and the other hot (cold) junction is on the bulk silicon frame. The temperature of the membrane is free to vary under an external source of heat (circulating air or radiation). Under appropriate conditions, the thermal gradient between the membrane and the silicon frame can reach 10~K over a distance of 150~$\mu$m. The thermoelectric voltage $V$ generated by the temperature gradient between the membrane and the silicon frame is collected by the two external contacts.} 
	\label{fig1}
\end{figure}

The concept at the basis of this work is twofold: first, to take advantage of the reduced thermal coupling between the suspended thermoelectric junction on the membrane and the heat bath (silicon frame) to favor heat exchange with the surroundings (air or radiation) and second to decrease the electrical resistance of the final device in order to use it as generator. It is made possible thanks to the elaboration of small suspended nanoTEG acting like individual nanosource of energy obtained by micro and nanostructuration (see Fig.~\ref{fig1}). These nanoTEGs can then be duplicated in series and/or parallel depending on the expected internal resistance; the microstructuration favoring at the same time the thermal insulation of the central part of the membrane and heat exchange through air or radiation. 

To optimize the efficiency of the device, we have used bismuth telluride alloys as thermoelectric materials for their high conversion efficiency close to room temperature \cite{bourgault2008,bourgault2018}. As their mechanical properties do not allow to make self- suspended structure, silicon nitride (widely used in MEMS industry) is used as a physical support taking profit of its mechanical stability and its low thermal conductivity. The use of large membrane and planar configuration boost surface dependent heat transfers like convection or radiation. The sensitive thermoelectric junctions are installed on the membrane and on the silicon frame to convert temperature gradients in TE voltage.

As compared to a regular macroscopic TE modules that are quite massive, this technical solution allows collecting energy from various small sources. In a regular TEG, the thermal conductance is given by the n and p TE pillars themselves. In a macroscopic module, the thermal conductance between the hot side and the cold side is given by the TE legs; it is of $10^{-4}$~W.K$^{-1}$ for typical leg dimensions of 1~mm$\times$1~mm$\times$5~mm. Here, since the nanTEGs are planar by construction, the thermal coupling of the sensitive part is only ensured by the TE thin films and the SiN suspending arms. Indeed, first, the aspect ratio of the suspending arms (small thickness, width and long length) is highly favorable for thermal insulation. Second, by using low thermal conductivity materials like amorphous silicon nitride and bismuth telluride, the thermal link to the heat bath is severely reduced as compared to the use of polysilicon \cite{perez2014,yuan2015,sakata2015}. This leads to thermal conductance of the order of $10^{-7}$~W.K$^{-1}$, few orders of magnitude smaller than their macroscopic counterparts. In other words, this very low thermal conductance and the very loss mass of the membrane allows to create a thermal gradient of one degree when the membrane exchange a tenth of picojoule with its environment. These characteristics and the very small mass of the thin silicon nitride membrane lead to significantly enhanced thermal gradients when air convection is present along with a very fast thermal time response.

\begin{figure}
	\begin{center}
		\includegraphics[width=10cm]{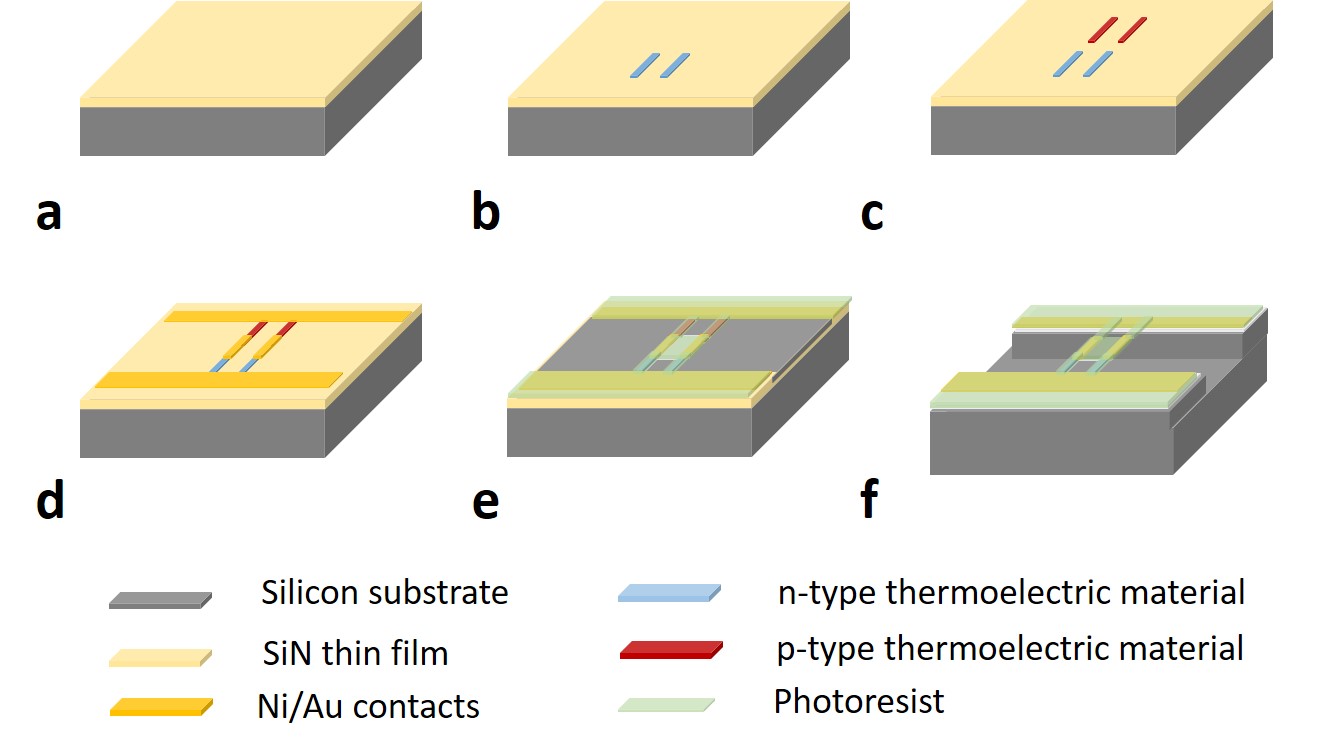}
	\end{center}
	\caption{Sketch of the main steps of the elaboration process of the nanoTEG devices : a) Si substrate with SiN top layer, b) structuration of the n-type Bi$_{2-x}$Te$_3$Se$_x$ thin film by lift-off process, c) idem for the p-type Bi$_{2-x}$Sb$_x$Te$_3$ thin film, d) and the Ni metallic contact, e) etching of the SiN top layer using SF$_{6}$ RIE etching, the electrical contact being protected by photoresist f) Suspension of the nanoTEGs with XeF$_{2}$ dry etching process.}
	\label{fig2}
\end{figure}

Let us first described the fabrication of an individual single nanoTEG cell. It is built by using two kinds of Bi$_{2}$XTe$_3$ thin TE films 300~nm thick (X being either antimony (Sb) for the p-type  Bi$_{2-x}$Sb$_x$Te$_3$ or selenium (Se) for the n-type is Bi$_{2-x}$Te$_3$Se$_x$). The TE thin films are deposited by reactive sputtering, structured by regular clean room processes and connected on-chip using a contact made of a metallic nickel thin film. As it can be seen in Fig.~\ref{fig1}, the membrane (100~nm thick) and the suspending arms (150~$\mu$m long, 6~$\mu$m wide) are made of SiN. The suspending arms are the mechanical support on which the TE thin films are deposited (150~$\mu$m long, 5~$\mu$m wide). After several microfabrication steps dedicated to the structuration of the TE junctions, membrane and supporting arms, via are opened into silicon nitride using SF$_{6}$ RIE etching. Membrane and arms are finally suspended using XeF$_{2}$ dry etching process. The main microfabrication steps are described in Fig.~\ref{fig2}. Then, by construction, if the silicon chip is installed on a hot surface, a temperature gradient will appear between the silicon frame and the suspended membrane, the temperature of the latter remaining close to the on of air thanks to convection. A thermoelectric voltage will then be generated across the TE junctions and collected on the two external contacts.

\begin{figure}
	\begin{center}
		\includegraphics[width=10cm]{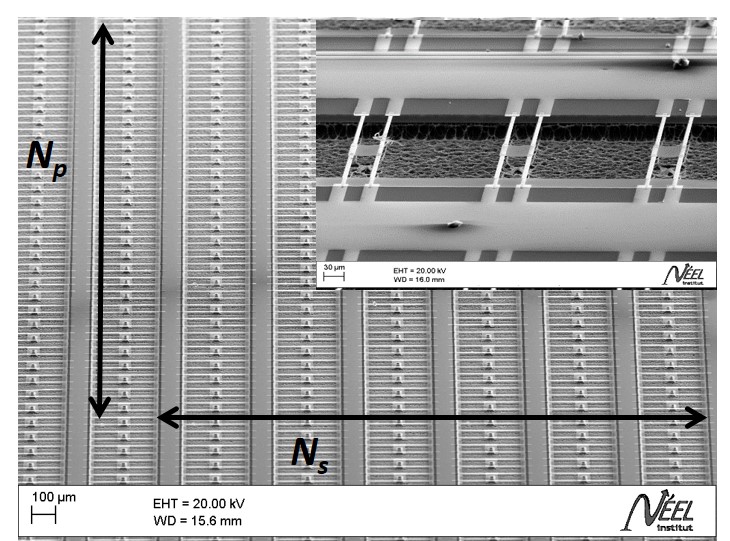}
	\end{center}
	\caption{SEM picture of a network of hundred of nanoTEG cells. The nanoTEGs are assembled in series and parallel for adapting the internal electrical resistance to the impedance of the electrical load. $N_{s}$ is the number of nanoTEG in series and $N_{p}$ the number of nanoTEG in parallel. In inset, a close-up view of three membranes of the network.}
	\label{fig3}
\end{figure}

The final nanoTEG module is then made out of thousands of identical single cells \cite{brevet1}. They are assembled in such a way (series and parallel) that the total electrical impedance can be significantly reduced if needed. Actually, this arrangement is very flexible and versatile, it can be adjusted to fit the targeted electrical resistance of the load. As pointed out by numerous works, the internal electrical resistance of the TE module is a point at least as important as thermoelectric performances. Reducing the internal resistance helps maximizing the power generated by the nanoTEG and permits its use with standard low power electronic components \cite{yuan2015}. An example of a large array of such nanoTEG cells is shown in Fig.~\ref{fig3}. Since the fabrication of these TE modules is made using microelectronic technologies, it can be up-scaled easily to make large surface of planar nanoTEGs.

\subsection*{Measure, characterization and performance.} 

More than ten different nanoTEG networks were fabricated and tested going from single cell TEG as shown in Fig.~\ref{fig1} to large array of membrane as shown in Fig.~\ref{fig3}. Several different physical properties are of great importance for a deep characterization of our nanoTEGs: the thermal conductance between the suspended structure and the heat bath, the resistance of the electrical contact at the junction between the p and n types TE materials, the Seebeck coefficient of the bismuth telluride and finally, the overall performance of the TEG to generate an electrical power harvested from its thermal environment.

\begin{table}
	\begin{center}
		\begin{tabular}{|c|c|c|c|c|c|}
			\hline
			Materials   & $e$  (nm)  & $T_{rec}$ (°C) & $\rho$ (m$\Omega$.cm) & $S$  ($\mu$V.K$^{-1}$) & $ S^2\sigma$ (Watt.m$^{-1}$.K$^{-2}$)  \\ \hline \hline
			Bi$_{2-x}$Sb$_x$Te$_3$ &  300    & 275 & 5  &  137  & $3.89 \times 10^{-4}$ \\ \hline
			Bi$_{2-x}$Te$_3$Se$_x$ &  300    & 175  & 2  &  -54  & $1.43 \times 10^{-4}$  \\ \hline
			Bi$_{2-x}$Sb$_x$Te$_3$ in \cite{bourgault2008}&  300    & 275 & 3  &  250 & $ 2.1 \times 10^{-3}$ \\ \hline
			Bi$_{2-x}$Te$_3$Se$_x$  in \cite{bourgault2008} &  300    & 250  & 1  &  -200  & $ 4.0 \times 10^{-3}$  \\ \hline
			
		\end{tabular}
	\end{center}
	\caption{Deposition parameters and thermoelectric properties of the two n and p-type bismuth telluride thin films. $e$ is the thickness of the TE materials, $T_{rec}$ is the temperature at which the TE materials are annealed, $\rho$ the resistivity of the TE film, $S$ the Seebeck coefficient measured at room temperature, and $S^2\sigma$ the power factor. The first two lines are related to parameters as measured on TE thin films in this work. As a comparison, the last two bottom lines of the table show the optimized parameters as obtained in previous studies made by Bourgault \textit{et al.} \cite{bourgault2008}. The Seebeck of the TE thin films used in this work are smaller than the accepted state-of-the-art value, showing the room of improvement that still exists for the nanoTEGs.}
	\label{table1}
\end{table}

First, the thermal properties of each constituting materials of the TEG have been measured independently. The thermal conductivity of SiN and Bi$_{2}$XTe$_3$ based TE materials has been measured using on membrane 3$\omega$ measurements \cite{sikora1,sikora2}. At room temperature, SiN exhibits a thermal conductivity below 3~W.m$^{-1}$.K$^{-1}$ and a value around 1~W.m$^{-1}$.K$^{-1}$ has been found for thin films of Bi$_{2}$Te$_3$ based materials \cite{sikora1,sikora2,ftouni}. This gives a thermal conductance of 10$^{-7}$~W.K$^{-1}$ for the four suspended beams showing that the membrane is highly thermally isolated from the frame. The electrical resistivity of each thermoelectric legs ($\rho$$_{p}$ and $\rho$$_{n}$) has been measured at room temperature at 5~m$\Omega$.cm for p-type and 2~m$\Omega$.cm for n-type using standard four probe experiment carried out on dedicated samples. Finally, the Seebeck coefficients have been measured using a home-made experimental setup for each composition of the Bi$_{2}$XTe$_3$ films on separated samples having the same geometry and contacts \cite{bourgault2008}. The TE thin films are not yet fully optimized because the Seebeck coefficient and the electrical conductivity that are summarized in Table~\ref{table1} are below the state-of-the-art value (see Table~\ref{table1}) \cite{bourgault2008,bourgault2018}. 
The internal resistance of the nanoTEG assembly $R_{sim}$ can be estimated from the relation: 
\begin{equation}
R_{sim} = N_s (\rho_{p} L+\rho_{n} L)/ (N_p w e)
\label{equation1}
\end{equation}

where $L$ is the length of the n and p-type arms, $w$ their width and $e$ their thickness; $N_{p}$ and $N_{s}$ being respectively the number of nanoTEGs in parallel and series. The expected resistance $R_{sim}$ for the studied array composed of 35 membranes in parallel and 12 membranes in series is 880~$\Omega$ (see Table~\ref{table2}). The differences observed between the simulated values from Eq.~\ref{equation1} and the measurements of the actual resistance of the four samples lies within 10\%. Intrinsic variation of resistance between samples can originate from inhomogeneity in the layer geometry, broken membrane or could be due to contact resistance. Regarding the contribution of the contact to the overall resistance, we can estimate that the contact resistance between metallic contact and Bi$_{2}$XTe$_3$ thin film to be below $10^{-6}$~$\Omega$.cm$^{2}$, a value in fair agreement with the state-of-the-art \cite{gupta}. This estimated value leads to contact resistances of several Ohms for a contact surface of few micrometer square; a resistance that will be negligible as compared to the resistance of the TE films themselves. After XeF$_2$ etching, the resistance of the whole device is increased by a factor of two. This increase has been observed for all the samples. If the true origin of that change is still under investigation, it may be assigned to the slight deterioration of the electrical properties of the TE thin films or the electrical contact to the nickel thin film.

\begin{table}
	\begin{center}
		\begin{tabular}{|c|c|c|c|c|}
			\hline 
			Samples    & S1   & S2  & S3  & 	 S4  \\ \hline  \hline 
		$R^{bf}$ ($\Omega$) &  1050    & 970    &  1030   &   910    \\ \hline 
		
		$R^{af}$  ($\Omega$) & 1980   &  2160 & 2590  &   2110     \\ \hline 
		
		\end{tabular}
	\end{center}
	\caption{Electrical characteristics of samples containing an array of 420~membranes over 0.5~cm$^2$. The resistance measurements have been performed before ($bf$) and after ($af$) the XeF$_2$ etching of Si for releasing the membranes. The XeF$_2$ etching fabrication step is influencing the value of the resistance of the Bi$_{2}$XTe$_3$ film or its contact resistance with the Ni.}
	\label{table2}
\end{table}

\begin{figure} [ht]
	\begin{center}
		\includegraphics[width=10cm]{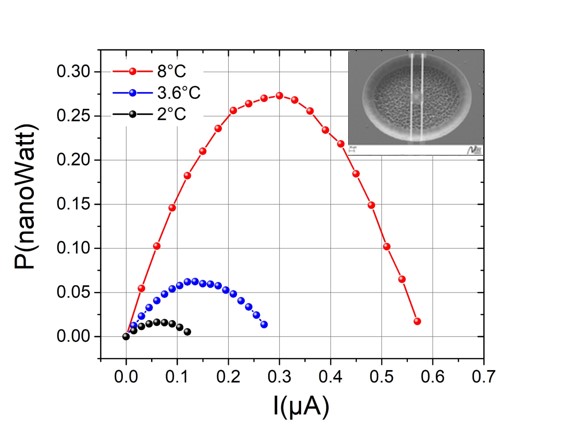}
	\end{center}
	\caption{Power generated by a single nanoTEG cell as a function of the bias current when the membrane is used in Peltier mode. Temperatures are determined from the thermoelectric voltage. The temperature differences between the membrane and the silicon frame used in the experiment are mentioned. In inset, the membrane on which the measurement has been done is shown.} 
	\label{fig4}
\end{figure}

\begin{figure}
	\begin{center}
		\includegraphics[width=10cm]{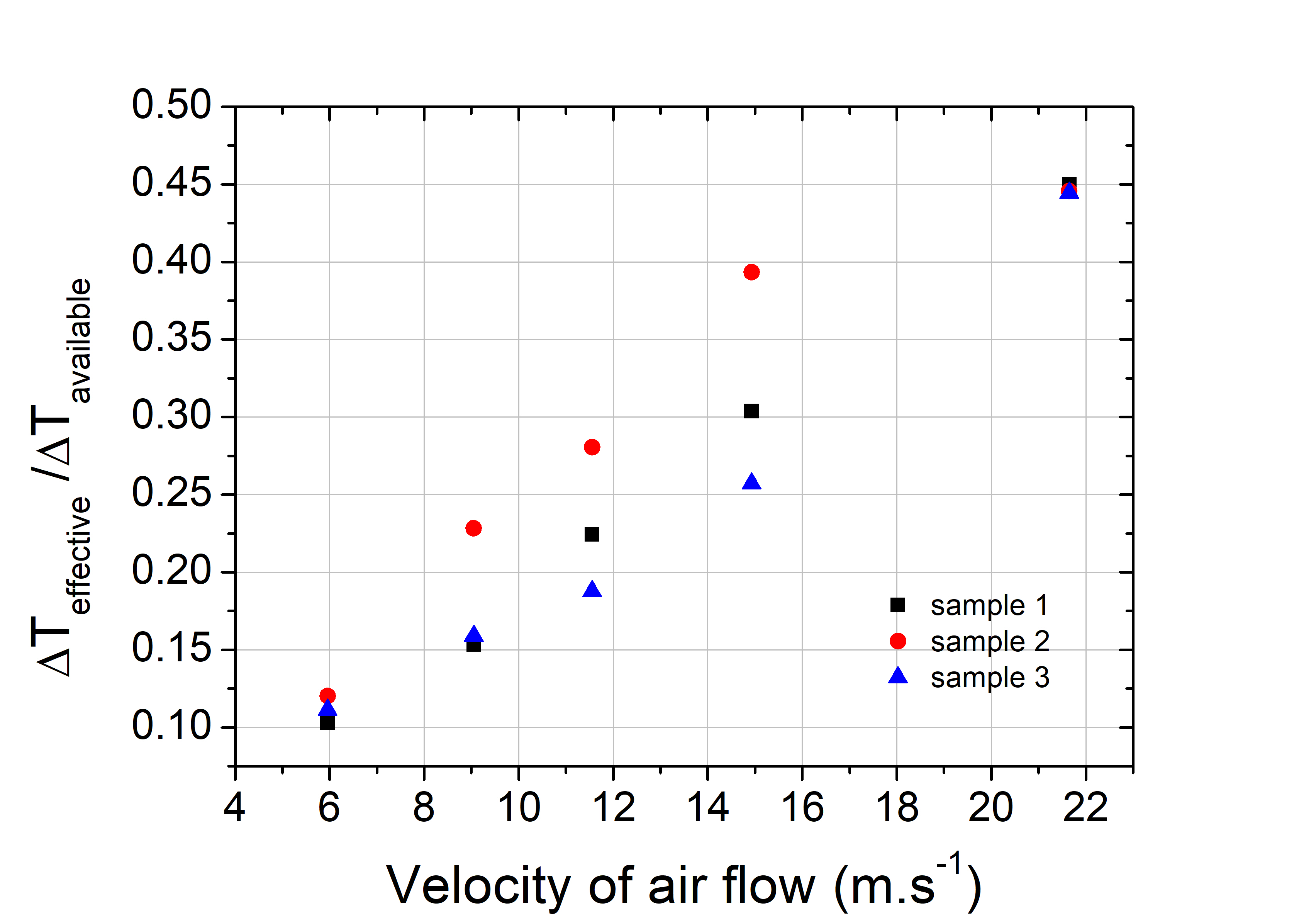}
	\end{center}
	\caption{Ratio between the actual effective temperature gradient as measured on the membranes and the temperature gradient available for different air-flow speeds. Results show that up to 60\% of the available temperature gradient is converted into a useful thermal gradient leading to the appearance of a 10~K temperature difference between the membrane and the silicon frame.} 
	\label{fig5}
\end{figure}

The performance of the nanoTEG has been measured on single membranes, as shown in Fig.~\ref{fig4}, as well as on a large network of membranes. A temperature gradient is imposed by the mean of a cold air flow blown on the chip carefully glued on a hot plate regulated initially at a temperature of 320~K. The cold junction located on the membrane is cooled down by the air flow while the silicon substrate coupled to the hot plate will serve as the hot heat sink. The temperature difference between the membrane and the Si frame is probed by measuring the voltage appearing across the TE junctions. When the air flow speed is increased the voltage measured at the end of the device increases. This shows that the nanoTEG device is very sensitive to forced convection. Fig.~\ref{fig5} shows the variation of the ratio between the effective temperature gradient as measured in the experiment and the available temperature gradient (the gradient between the temperature of the hot plate and the temperature of the air flow). It is worth noticing that this ratio reaches 60~percent for airflow speed of 25 m.s$^{-1}$. 

\begin{figure}
	\begin{center}
		\includegraphics[width=10cm]{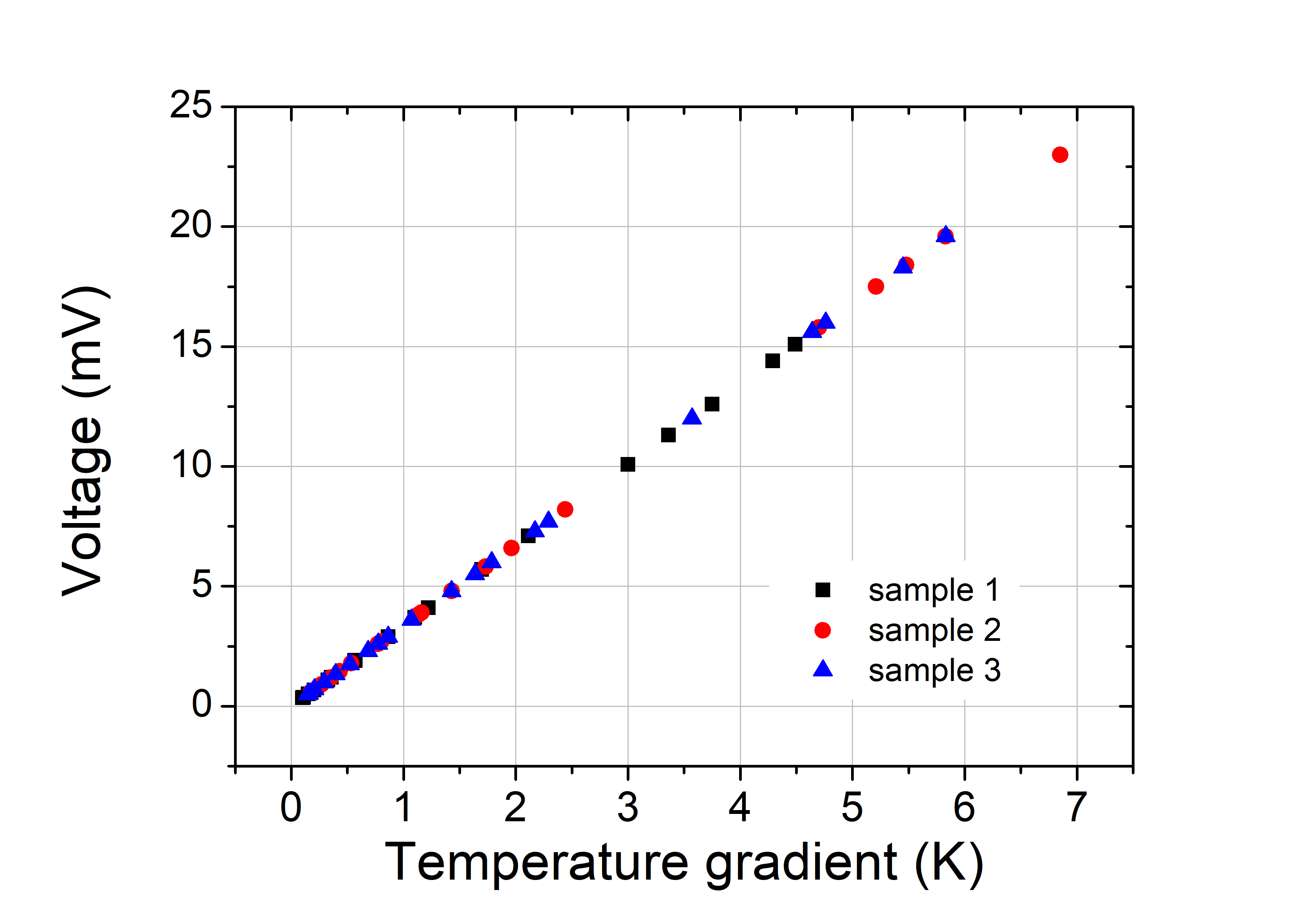}
	\end{center}
	\caption{Variation of the thermoelectric voltage as a function of the effective thermal gradient for three different samples. As expected, the variation is linear.} 
	\label{fig6}
\end{figure}

These conditions of air flowing are frequently found in moving parts that necessitate on-board sensor application. The power generated by the nanoTEGs has been measured for different temperature gradients. Results are displayed in Fig.~\ref{fig6}; it shows the expected linear variation of the generated thermoelectric voltage as function of the effective temperature gradient. The power output generated by an assembly of 420~membranes covering 0.5~cm$^2$ surface subjected to a temperature gradient of 10~K has been measured to be of 0.15~$\mu$Watt, giving for 1~cm$^2$, a power of 0.3~$\mu$Watt.

This result is significant since it is obtained with not-fully optimized thermoelectric materials as shown in Table~\ref{table2}. Indeed, by considering the standard TE properties of bismuth-telluride alloys a power output of more than 10 times higher can be generated, meaning 3~$\mu$Watt for a 10~K temperature gradient over a 1~cm$^2$ surface. Improvement in the technological process will permit increasing the membrane density by a factor of, at least, three. By working with larger surface (10~cm$^2$) along with larger temperature gradient, the power generated will be sufficient to feed a low energy communication protocol and exchange information on temperature or airflow speed.

As a concluding remarks, it can be mentioned that such miniaturized planar suspended thermoelectric device could also be used for cooling applications, and not only for energy harvesting. For example, small systems located at the center of the suspended membranes can be cooled down by current injection in the nanoTEGs device, using such device in Peltier mode. 

\subsection*{Conclusion} 

Miniaturiazed planar suspended thermoelectric devices have been developed for low power energy harvesting. This device is designed to convert small intermittent gradient of temperature into electrical power by using thermally isolated membranes as heating or cooling platform. The geometry of the device permits to adapt the resistance of the thermogenerator for each particular use case. The efficiency of the nanoTEG is sufficient to expect its application in real wireless sensor network. By construction, the architecture of the device allows the appearance of high thermal gradients over very small distance (60~K.mm$^{-1}$). Using optimized TE materials, more than few tens of microWatt can be generated hence permitting a wireless communication between the sensor and the network.

Regular microelectronic fabrication processes are used to elaborate such nanoTEG modules, meaning that the production can be easily upscaled at relatively low cost along with a high integrability. Finally, it has to be stressed that better performances and evolution of the technology are foreseen. Denser array of membranes, optimized new geometry and higher Seebeck coefficients can be easily obtained boosting the performances of the nanoTEGs towards higher generated power. This shows that membrane based thermoelectric energy harvesters are very promising building blocks for applications where hot or cold air is present offering excellent perspectives for autonomous and connected physical sensing.

\subsection*{Acknowledgments} 

We thank the micro and nanofabrication facilities of Institut N\'eel CNRS: the Pole Capteurs Thermom\'etriques et Calorim\'etrie (E. Andr\'e, G. Moiroux and J.-L. Garden) and Nanofab (L. Abbassi, B. Fernandez, T. Fournier, G. Juli\'e and J.-F. Motte) for their advices in the preparation of the samples. The authors acknowledge the financial support from EU MERGING Grant No. 309150, the MODULO project Premat CNRS program in 2015-2016, and the SATT Linksium financial support for maturation of the Mo\"{\i}z project in 2017 (https://moiz-eh.com/).

\end{document}